\documentclass[aps,nofootinbib,amsfonts,groupedaddress]{revtex4}
\usepackage{amsmath}
\usepackage{epsfig,epsf}
\usepackage{graphicx}
\usepackage{verbatim}
\usepackage{epstopdf}

\def\beq{\begin{equation}}
\def\eeq{\end{equation}}
\def\bea{\begin{eqnarray}}
\def\eea{\end{eqnarray}}
\def\eq#1{{Eq.~(\ref{#1})}}

\newcommand{\GV}{\mbox{GeV}}
\newcommand{\bas}{\bar{\alpha}_s}
\newcommand{\as}{\alpha_s}
\newcommand{\Lb}{\left(}
\newcommand{\Rb}{\right)}
\newcommand{\nn}{\nonumber}

\begin{document}

\voffset1.5cm
\title{Gluon saturation and inclusive hadron production at LHC }
\author{Eugene Levin$^{1,2}$ and Amir H. Rezaeian$^1$}
\affiliation{
$^1$ Departamento de F\'\i sica, Universidad T\'ecnica
Federico Santa Mar\'\i a, Avda. Espa\~na 1680,
Casilla 110-V,  Valparaiso, Chile \\
$^3$ Department of Particle Physics,  Tel Aviv University, Tel Aviv 69978, Israel}
 
\date{\today}
\begin{abstract}
In high density QCD the hadron production stems from decay of
mini-jets that have the transverse momenta of the order of the
saturation scale. It is shown in this paper that this idea is able to
describe in a unique fashion both the inclusive hadron production for
$\sqrt{s} \geq\,546~\GV$ including the first data from LHC
and the deep inelastic scattering at HERA. Recently reported data from
ALICE, CMS and ATLAS including inclusive charged-hadron
transverse-momentum and multiplicity distribution in $pp$ collisions
are well described in our approach. We provide predictions for the
upcoming LHC measurements.
\end{abstract}
\maketitle


\section{Introduction}

The first LHC data \cite{AL1,CMS,ATLAS,ALICE} on inclusive hadron
production call for a theoretical understanding of these processes
based on QCD.  At first sight the inclusive hadron production is a
typical process that occurs at long distances where one has to use the
non-perturbative methods of QCD. Therefore, the field of long distance
processes seems to be a relevant subject to the domain of
high-energy phenomenology with the main ingredients: soft Pomeron and
secondary Reggeons. Such phenomenology is able to describe inclusive
hadron production data (see Ref.~\cite{GLMINC} and references therein)
but cannot be considered satisfactory since both soft Pomerons and
Reggeons cannot be explained in terms of QCD ingredients; quarks and gluons. It should be also mentioned that
the increase with energy of the average transverse momentum of the
produced hadron observed experimentally \cite{CMS,ATLAS} cannot be explained
in the Reggeon approach.

However, high density QCD \cite{GLR,MUQI,MV,B,MUCD,K,JIMWLK} leads to
a completely different picture of inclusive hadron production. In this
approach the system of parton (gluons) at high energy forms a new
state of matter: Color Glass Condensate (CGC). In the CGC picture, at
high energy the density of partons $\rho_p$ with the typical
transverse momenta less than $Q_s$ reaches a maximum value, $\rho_p
\propto 1/\as \,\,\gg\,\,1$ ($\as$ is the strong coupling
constant). $Q_s$ is the new momentum scale (saturation momentum) that
increases with energy. At high energies/small Bjorken-x, $ Q_s
\,\,\gg\,\,\mu$ where $\mu$ is the scale of soft
interaction. Therefore, $\as\Lb Q_s\Rb \,\,\ll\,\,1$ and this fact
allows us to treat this system on solid theoretical basis. On the
other hand, even though the strong coupling $\alpha_s$ becomes small
due to the high density of partons, saturation effects, the fields
interact strongly because of the classical coherence. This leads to a
new regime of QCD with non-linear features which cannot be investigated
in a more traditional perturbative approach.
 
In the framework of the CGC approach the secondary hadrons are originated
from the decay of gluon mini-jets with the transverse momentum equal
to the saturation scale $Q_s(x)$. The first stage of this process is
under theoretical control and determines the main characteristics
of the hadron production, especially as far as energy, rapidity and
transverse momentum dependence are concerned. The jet decay,
unfortunately, could be treated mostly phenomenologically.  However,
we can hope that the phenomenological uncertainties would be reduced
to several constants whose values will be extracted from the
experiment.

Actually, such a description has passed the first check with the
experimental data: the KLN paper \cite{KLN} explains the main
features of inclusive hadron production in heavy ion-ion and
hadron-ion  as well as proton-proton collisions \cite{KLNLHC} at
RHIC. In this paper we wish to improve the KLN approach by
introducing two new elements: the probability to find gluon with fixed
transverse momentum that describes the deep inelastic scattering (DIS)
data and that satisfies the Balitsky-Kovchegov
\cite{B,K} non-linear equation; and a different description of inclusive hadron production at low
transverse momenta of gluons. Over all success of our description
indicates universality of the saturation physics which can be further
tested at LHC and future collider experiment.  

In the next section we discuss the $k_t$-factorization and main formulas that we use. In
particular, we consider the interrelation between the color dipole
scattering amplitude and the unintegrated gluon density that follows
from the recent development of high density QCD \cite{KTINC}. An
important improvement here to the previous works based on the KLN
approach is the explicit inclusion of the impact-parameter dependence
of the saturation scale. Section III is devoted to comparison with the
experimental data and to discussion of various predictions for higher LHC
energies. As a conclusion, in Sec. IV we highlight our main results and predictions for LHC.


\section{Inclusive gluon production in high density QCD}
The gluon jet production in hadron-hadron collisions can be described by $k_t$-factorization given by \cite{KTINC}, 
\beq \label{MF1}
\frac{d \sigma}{d y \,d^2 p_{T}}\,\,=\,\,\frac{2\alpha_s}{C_F}\frac{1}{p^2_T}\int d^2 \vec k_{T}\,\,\phi^{h_1}_G\Lb x_1;\vec{k}_T\Rb\,\phi^{h_2}_G\Lb x_2;\vec{p}_T -\vec{k}_T\Rb, 
\eeq
where $x_{1,2}=(p_T/\sqrt{s})e^{\pm y}$, and $p_T$ and $y$ is the transverse
momentum and rapidity of the produced gluon jet. $\phi^{h_i}_G$ are the probability to find a gluon that carries
$x_i$ fraction of energy with $k_T$ transverse momentum and
$C_F=(N^2_c-1)/2N_c$ is the $SU(N_c)$ Casimir operator in the
fundamental representation with the number of colors equals $N_c$.

For a proof of $k_t$-factorization see Ref. \cite{KTINC} and also
Refs. \cite{BRINC,CMINC,KLINC,LPINC,KLPINC} which confirm the former
proof\footnote{Ref.~\cite{BSVINC} states that \eq{MF1} is not
correct. Unfortunately, there is no discussions in the paper why their
result is so different from the other published papers. However, Braun
has recently shown that Ref. \cite{BSVINC} actually leads to the
$k_t$-factorization \cite{yuri-b}.}.  We need to recall that the proof for the
$k_t$-factorization was given for the scattering of a diluted system
of partons, say for virtual photon, with a dense one. Our main idea is
that we have gluon saturation for proton-proton scattering or in other
words, we are dealing with interactions of two dense systems of
partons (gluons). Therefore, the $k_t$-factorization has to be
considered here as an assumption. It should be noticed that the
proof given in Refs.~\cite{KTINC,BRINC,CMINC,KLINC,LPINC,KLPINC} shows that the $k_t$-factorization is
valid in the situation where two scales of hardness: the transverse
momentum of the produced gluon ($p_T$) and the saturation scale are both larger than the scale of the soft interaction
($\mu$).  For dense-dense system scattering we have actually three
scales: $p_T$ and two saturation scales. However, only for the
kinematic region where both $x_1$ and $x_2$ are small and for $p_T$ which is
smaller than both saturation scales we have to make an assumption
about $k_t$-factorization. In other cases that one of the saturation scales
is small, we are dealing with diluted-dense system scattering. We believe that the
$k_T$-factorization is currently the best tools at our disposal for
the processes considered in this paper.

The unintegrated gluon density $\phi^{h_i}_G\Lb x_1;\vec{k}_T\Rb$ and
color dipole-proton forward scattering amplitude $N\Lb x_i, r_T; b
\Rb$ are related in a very specific way \cite{KTINC}. This relation
reads as follows
\beq \label{MF2}
\phi^{h_i}_G\Lb x_i;\vec{k}_T\Rb\,\,=\,\,\frac{1}{\alpha_s} \frac{C_{F}}{(2 \pi)^3}\,\int d^2 \vec b \,d^2 \vec r_T
e^{i \vec{k}_T\cdot \vec{r}_T}\,\,\nabla^2_T\,N^{h_i}_G\Lb y_i = \ln(1/x_i); r_T; b \Rb,
\eeq
with 
\beq \label{MF3}
N^{h_i}_G\Lb y_i = \ln(1/x_i); r_T; b \Rb\,\,=\,\,2 \,N\Lb y_i = \ln(1/x_i); r_T; b \Rb\,\,-\,\,N^2\Lb y_i = \ln(1/x_i); r_T; b \Rb,
\eeq
where $N^{h_i}_G\Lb y_i = \ln(1/x_i); r_T; b \Rb$ is the dipole-hadron ($h_i$) forward scattering amplitude which satisfies the Balitsky-Kovchegov equation.
In the above, $r_T$ denotes the transverse dipole size and $\vec b$ is the impact parameter of the scattering. 

\eq{MF3} looks very natural at large $N_c$. Indeed, for the color dipole amplitude in 
the Glauber form $ N = 1 -\exp\Lb - \Omega/2\Rb$ ($\Omega$ is the opacity), equation  \eq{MF3} leads to $N_G =1 - \exp\Lb -
\Omega\Rb$ as it should be for the scattering of the two dipoles of
the same sizes. We recall that a colorless gluon-probe just creates such two quark-antiquark dipoles, and the $N_G$ is directly
related to the gluon density.

\begin{figure}[!t]
              \includegraphics[width=8 cm] {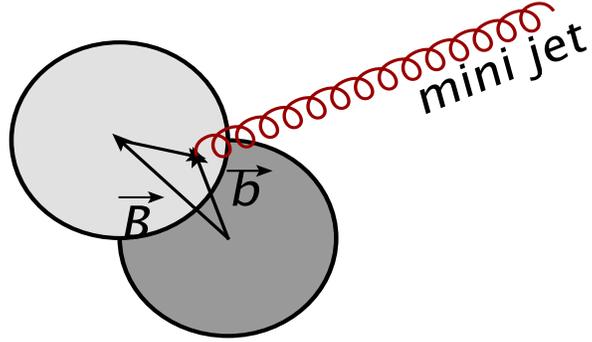}
   \caption{Mini-jet production in hadron-hadron collisions in the transverse plane. The impact-parameter between two hadrons is $\vec B$. }
\label{f0}         
\end{figure}

Substituting  \eq{MF2} in \eq{MF1}, and after analytically performing some integrals, we obtain \cite{KTINC}
\bea \label{MF4}
\frac{d \sigma}{d y \,d^2 p_{T}}\,\,& & \frac{2C_F}{\alpha_s (2\pi)^4}\,\frac{1}{p^2_T}\int d^2 \vec b \,d^2 \vec B \,d^2 \vec r_T\,e^{i \vec{k}_T\cdot \vec{r}_T}\,\,\nabla^2_T\,N^{h_1}_G\Lb y_1 = \ln(1/x_1); r_T; b \Rb\,\,\nabla^2_T\,N^{h_2}_G\Lb y_2 = \ln(1/x_2); r_T; |\vec b-\vec B| \Rb. \nonumber\\
\eea
In the above equation, $\vec B$
is the impact parameter between center of two hadrons and $\vec b$ is the impact
parameter of the produced mini-jet  from the center of the hadron, see Fig.~\ref{f0}.

\subsection{Choice of color dipole scattering amplitude }
As it can be seen from Eq~(\ref{MF2},\ref{MF4}), we need here an
impact-parameter dependent color-dipole forward amplitude. We will
show later that the inclusion of the impact-parameter is very
important in our approach and should not be ignored.  The
dipole-proton forward scattering amplitude $N\Lb Y; r;b\Rb$ (with $Y= \ln(1/x)$) can be in
principle found by solving the perturbative nonlinear small-$x$
Balitsky-Kovchegov (BK) \cite{B,K} or
Jalilian-Marian-Iancu-McLerran-Weigert-Leonidov-Kovner (JIMWLK) \cite{JIMWLK}
quantum evolution equations.  Unfortunately, numerical solution to
these non-linear equations in the presence of the impact-parameter is
very challenging \cite{bk-b} and is not yet available.
Moreover, a numerical solution does not give us the full control on the
phenomenological parameters that have been used and we certainly lose
the transparency and simplicity of physical interpretation if we rely
only on the numerical solutions. Therefore, we choose a different
approach to the solution of the BK equation that was
suggested in Ref.~\cite{IIMDIS}. First, we recall that the BK equation
predicts the geometric scaling behavior \cite{GS}, namely the amplitude $N\Lb Y;r;b\Rb$ is not a function of three variables but it is a
function of only one variable $\mathcal{Z}^2=\,r^2\,Q^2_s(x; b)$ ( $N\Lb Y;r;b\Rb\,=\,F\Lb \mathcal{Z}\Rb$)
where $Q_s(x; b)$ is the saturation momentum\footnote{Notice that here we assumed that the geometric scaling is also
valid in the presence of impact-parameter dependence of the saturation scale.
It should be stressed that the proof of the geometric scaling
behavior \cite{GS} could be easily generalized to the case of the
scattering amplitude that depends on the impact-parameter $b$.  In
the analytical solution of Ref.~\cite{LTSOL} which gives the theoretical
 basis for the chosen parametrization of the dipole-amplitude here, the $b$-dependence is taken into
account, therefore, this solution gives a theoretical example of
the general proof.}. We also 
know \cite{LTSOL} the behavior of the scattering amplitude deeply in
the saturation region ($\mathcal{Z}\,\gg\,1$) 

\beq \label{CA1}
N\Lb Y; r; b\Rb\,\,=\,\,\,1 \,\,-\,\,\exp\Lb -\,\frac{\chi(\gamma_{cr})}{2 ( 1 - \gamma_{cr})}\,\ln^2 \mathcal{Z}\Rb, 
\eeq
where $\chi(\gamma)$ is the BFKL kernel
\beq \label{CA2}
\omega\Lb \gamma\Rb\,=\,\,\,\bas \chi\Lb \gamma\Rb\,\,=\,\,\bas \left\{2 \psi(1)\,-\,\psi(\gamma))\,-\,
\psi(1 - \gamma)\right\},
\eeq
with a notation  $\bas=\alpha_s N_c/\pi$. In above, we define $\psi(x) = d \ln \Gamma(x)/d x$ and $\Gamma(x) $ is the Euler function. 
The parameter $\gamma_{cr}$ is the solution to the following equation
\beq \label{CA3}
\frac{d \chi\Lb \gamma_{cr}\Rb}{d \gamma_{cr}}\,\,=\,\,-\,\frac{\chi\Lb \gamma_{cr}\Rb}{ 1\,-\,\gamma_{cr}}.
\eeq
In Ref. \cite{LTSOL} a solution was found for the entire kinematic region for a simplified BFKL kernel, namely,
instead of \eq{CA2}, the following kernel was used, 

\bea \label{CA4}
\omega\Lb \gamma\Rb\,\,=\,\,\bas \left\{\begin{array}{l}\,\,\,\frac{1}{\gamma}\,\,\,\,\,\mbox{for}\,\,\, \mathcal{Z}\,=\,r Q_s(x)\,\leq \,1\,;\\ \\
\,\,\,\frac{1}{1 \,-\,\gamma}\,\,\,\,\,\mbox{for}\,\,\, \mathcal{Z}\,=\,r Q_s(x)\,>\,1\,; \end{array}
\right.
\eea
which describes only leading twist contribution to the full BFKL
kernel of \eq{CA2}.  The lesson from this solution is very
instructive: for $r^2Q^2_s(x)\,\leq\,1$, the amplitude $N$ satisfies the DGLAP (BFKL)
linear evolution equation with the boundary condition $N( Y;r;b) \,=
\,N_0\,=\, \text{Constant}$ for $r^2 = 1/Q^2_s(x)$ while for $r^2Q^2_s(x)\,>\,1$ we
have solution that has the form of \eq{CA1}. Using these general
features of the solution we choose the model suggested in
Ref. \cite{WAKO} which improves the earlier studies on this line \cite{IIMDIS, WAMOKO}. In this model the color dipole-proton
forward scattering amplitude is given by
\bea \label{CA5}
N\Lb Y; r; b\Rb\,\,=\,\, \left\{\begin{array}{l}\,\,\,N_0\,\Lb \frac{\mathcal{Z} }{2}\Rb^{2 (\gamma_s\,\,+\,\,\frac{1}{\kappa \lambda Y}\ln\Lb\frac{2}{\mathcal{Z} }\Rb)}\,\,\,\,\,\mbox{for}\,\,\mathcal{Z}\,=\,r Q_s(x)\,\leq\,2\,;\\ \\
1\,\,-\,\,\exp\Lb -A \ln^2\Lb B \mathcal{Z}\Rb\Rb\,\,\,\,\,\,\,\,\,\,\mbox{for}\,\,\mathcal{Z}\,=\,r Q_s(x)\,>\,2\,;\end{array}
\right.
\eea
where the saturation scale $Q_s(x;b)$ (denoted by $Q_s(x)$ for brevity) is given by
\beq \label{CA6}
Q_s(x;b)\,\,=\,\,\Lb \frac{x_0}{x}\Rb^{\frac{\lambda}{2}}\,\exp\left\{- \frac{b^2}{4 (1 - \gamma_{cr}) B_{CGC}}\right\}.
\eeq

As we have already mentioned \eq{CA5} as well as \eq{CA6} has the form
of the solution to the BK equation at a fixed QCD coupling. For $\mathcal{Z} <
1$ the effective anomalous dimension $
\gamma_{s}\,\,+\,\,\frac{1}{\kappa \lambda Y}\ln\Lb\frac{2}{\mathcal{Z}}\Rb$
with $\gamma_s = 1 - \gamma_{cr} $ follows from the BFKL (and
DGLAP) equation in the vicinity of the saturation line (see
Ref.~\cite{IIMDIS} for the detailed derivation).

For the leading order BFKL kernel with frozen QCD coupling the parameters of  \eq{CA5} and \eq{CA6} have the following values
\beq\label{CA7}
1 - \gamma_{cr}\,=\,0.63; \,\,\,\,\,
\lambda\,\,=\,\,\bas \frac{\chi\Lb \gamma_{cr}\Rb}{ 1\,-\,\gamma_{cr}}\,=\,4.88 \bas; \,\,\,\,\,
\kappa\,\,=\,\, \frac{\chi\prime\prime´\Lb \gamma_{cr}\Rb}{\chi\prime\Lb \gamma_{cr}\Rb }\,=\,9.9.
\eeq

The parameters $A$ and $B$ can be found from matching of $N$ and its logarithmic derivatives at $\mathcal{Z}=2$ while $N_0$ and $B_{CGC}$ remain fitting parameters.

Generally speaking, the model given by Eqs.~(\ref{CA5},\ref{CA6}) can be viewed as
an approximation to the solution of the BK equation. However, because
the $b$-dependent numerical solution to the BK equation is not yet
available \cite{bk-b}, we are doomed to resort to such an approximation. This
model differs from other saturation models on the market since it
apparently incorporates all known properties of the exact solution to
the BK equation including the $b$-dependence of the scattering
amplitude (see Ref.~\cite{LTSOL}).

The advantage of Eqs.~(\ref{CA5},\ref{CA6}) is that these equations give the possibility
to take into account the next-to-leading order (NLO) corrections.
Two features of the non-linear
low-$x$ equations can be calculated in the next-to-leading order
using the kernel of the linear equation: the energy behavior of the
saturation scale  \cite{GLR,MUTR,MUPE} and the behavior of the solution deeply in
the saturation domain \cite{LTSOL}.  It has been shown
that the NLO correction to the BFKL equation (and therefore BK equation)
are large and it changes considerably the value of $\lambda$ from
$\lambda \approx 0.9 $ to $ \lambda \approx 0.3$ for $\bas = 0.2$
\cite{TRI,DGSATMO}. The value of $\gamma_s$ in \eq{CA5} is also
affected by the NLO corrections as well as by the running QCD coupling
\cite{TRI,DGSATMO,ALKO}. 
It is therefore generally believed that
the higher order corrections to the NLO BK equation should be
important. The actual calculation of higher-order corrections to these
non-linear evolution equations still remains as a challenge. Since the
general behavior of the amplitude Eq. (9) will remain unchanged after
inclusion of higher-order corrections, we effectively incorporate the
higher-order corrections by taking the value of parameters $ \lambda,
\gamma_s, N_0$ and $B_{CGC}$ obtained from a fit to the DIS data at
low Bjorken-$x$ $x<0.01$ \cite{WAKO}. Therefore, the saturation model that
we use here gives also a good description of the HERA data at
low-$x$. In
order to simulate the behavior of gluon density at large $x\to 1$, we
product the unintegrated gluon density with $(1-x)^4$ as prescribed by quark counting rules \cite{bigx}.
This factor stems from the correct description of the HERA data on DIS.

\subsection{Physical observables}
The rapidity distributions of the mini-jets can be calculated using \eq{MF1}, 
\beq \label{PO1}
\frac{d N_{\mbox{mini-jet}}}{d \eta}\,\,=\,\,h[\eta]\frac{1}{\sigma_{nsd}}\int d^2 p_T \,\frac{d \sigma}{d y \,d^2 p_{T}}\left[\eq{MF1}\right],
\eeq
where $\eta$ is the pseudorapidity and $h[\eta]$ is the Jacobian
which takes account of the difference between rapidity $y$ and the
measured pseudo-rapidity $\eta$ \cite{KLN},
\beq \label{PO3}
h\Lb \eta,p_T\Rb \,\,= \frac{\cosh\eta}{\sqrt{\frac{m^2_{jet}  +  p^2_T}{ p^2_T}\  \,+\,\sinh^2\eta}},
\eeq 
where $m_{jet}$ is the mass of mini-jet. One also has to express rapidity
$y$ in \eq{MF1} in terms of pseudo-rapidity $\eta$. This relation is
given by

\beq \label{PO2}
y\Lb \eta, p_T\Rb \,\,=
\,\,\frac{1}{2} \ln\left\{\frac{\sqrt{\frac{ m^2_{jet}+  p^2_T}{ p_T^2} \,+\,\sinh^2\eta}\,\,+\,\,\sin\eta}{\sqrt{\frac{ m^2_{jet}  +  p_T^2}{p^2_T}\,+\,\sinh^2\eta}\,\,-\,\,\sinh \eta}\right\}. 
  \eeq

  The distribution Eq.~(\ref{MF1}) refers to the radiated gluons with zero
mass while what is actually measured experimentally is the
distribution of final hadrons.  We therefore should make an assumption
about hadronization of gluons which is entirely non-perturbative
process that has to be modeled in any approach due to lack of
understanding of the confinement of quarks and gluon in QCD. However,
it is well-known that the general assumption about hadronization leads
to the appearance of mass of the mini-jet which is approximately
equal to $m^2_{jet} \simeq 2 \mu p_T$ (see Ref.~\cite{KLN}) where $\mu$ is
the scale of soft interaction. The mini-jet mass $m_{jet}$ effectively
incorporates the non-perturbative soft pre-hadronization in the 
pseudo-rapidity space.  Accordingly, one should also correct the
kinematics every where in Eq.~(\ref{MF1}) due to the presence of a non-zero
mini-jet mass, namely replacing $p_T\to
\sqrt{p^2_T+ m^2_{jet}}$ in $x_1, x_2$ and also in the denominator of
$1/p_T^2$.  One can see that Eq.~(\ref{MF1}) has infrared divergence at
$p_T \to 0$ for the kinematic region $k_T \,\gg \,p_T$ when
$m_{jet}=0$. In Ref.~ \cite{KLN} it was suggested to integrate over $k_T
\,\leq
\,p_T$. The reason is that such an integration reproduces the
factorization formula at large $p_T \,\gg \,\mu$ for the DGLAP
evolution. However, as we explained above it is more natural to
replace $p_T$ by $\sqrt{p^2_T \,+\,m^2_{jet}}$ in Eq.~(\ref{MF1}) which
consequently also regulates the denominator due to the presence of a
non-zero mini-jet mass (the appearance of such mass is the
general property of the hadronization processes).

In \eq{PO1} we do not take into account the fragmentation of the
produced gluon (mini-jet) into hadrons. We rely on the principle of
Local Parton-Hadron Duality (LPHD) \cite{LPHD,LPHD1} namely the form of the
rapidity distribution will not be distorted by the jet decay and only
a numerical factor will differ the mini-jet spectrum from the hadron
one. We believe that it is better to use the LPHD scheme than to
deal with the fragmentation's functions for which we have no
theoretical justifications at low $p_T$.  It should be stressed that the same idea
has been used in the KLN approach which describes the rapidity
distribution of heavy-ion collisions data in a wide range of energies. This idea has also worked perfectly in $e^+ e^- $ annihilation into hadrons
\cite{LPHD,LPHD1}. 

We should stress that the value of inelastic non-singlet
diffractive (NSD) cross-section $\sigma_{nsd}$ cannot be calculated in our
approach and has  to be taken from the soft interaction models such
as in Refs. \cite{GLMM,KMR}. The NSD cross-section $\sigma_{nsd}$ is defined as
$\sigma_{nsd} = \sigma_{tot} - \sigma_{el} -
\sigma_{sd} - \sigma_{dd}$ where $\sigma_{el}$, $\sigma_{sd}$ and $\sigma_{dd}$ are the cross sections of elastic, single and double diffraction, respectively. 
However, the experimental data on $\sigma_{dd}$ is very limited \cite{dd-c}, $\sigma_{sd}$ is measured with rather large errors \cite{sd-c,tt-c}
and even for the total cross-section $\sigma_{tot}$ \cite{tt-c} we have two values at the Tevatron
energies \cite{tev-c}.  Therefore, we should stress that in this way we can only predict 
$d\sigma/d y$ rather than $dN_{ch}/dy$.  In order to overcome this problem, here we choose a different strategy:
the physical meaning of $\sigma_{nsd}$ in \eq{PO1} is the area of
interaction which can be calculated in our approach. Indeed,
using \eq{MF4} one can calculate the average impact parameter
for the inclusive production of the mini-jet 
\bea \label{PO5}
&&\Big \langle \vec b^2_{jet} \Big\rangle\,\, =\,\, \\
&&\frac{\int\frac{ d^2 p_T }{p^2_T}\int  d^2 \vec b \,d^2 \vec B \, d^2 r_T \Lb b^2 \,+\,|\vec b-\vec B|^2  \Rb\,e^{i \vec{k}_T\cdot \vec{r}_T}\,\,\nabla^2_T\,N^{h_1}_G\Lb y_1 = \ln(1/x_1); r_T; b \Rb\,\,\nabla^2_T\,N^{h_2}_G\Lb y_2 = \ln(1/x_2); r_T; |\vec b-\vec B|\Rb}{\int \frac{d^2 p_T }{p^2_T}\int \,d^2 \vec b \,d^2 \vec B \,d^2 r_T\,e^{i \vec{k}_T\cdot \vec{r}_T}\,\,\nabla^2_T\,N^{h_1}_G\Lb y_1 = \ln(1/x_1); r_T; b \Rb\,\,\nabla^2_T\,N^{h_2}_G\Lb y_2 = \ln(1/x_2); r_T; |\vec b-\vec B|\Rb}\nn.
\eea
The NSD cross-section  $\sigma_{nsd}$ is then equal to the average interaction area upto
a constant $\sigma_{NSD}=M \pi \Big \langle \vec b^2_{jet}
\Big\rangle$. The pre-factor $M$ will be determined and discussed later. 
We should draw the reader attention that such a picture for the
inelastic cross-section corresponds, in a sense, to the
geometric-scaling behavior of the scattering amplitude.  Indeed, the
high-density QCD deals with the partonic wave-function of a fast
hadron which describes a coherent system of partons (quarks and
gluon). At high energy the coherence of partons is destroyed during a
short time, and the partons, distributed as in the wave function, are
produced. These partons contribute to the inelastic cross-section.
The elastic (diffractive ) cross-section corresponds to a rare event
where the target does not destroy (or destroyed only partially) the
coherence of the gluons in the wave-function (see for example
Ref.~\cite{Geo-BB}). The geometric-scaling behavior as well as
the saturation phenomenon, in general, means that partons are
distributed uniformly in the transverse plane in the wave-function of
a fast hadron in a such way that the wave-function generates a
uniform distribution of the produced partons after the interaction with
the target. Therefore, the NSD (inelastic) cross-section is proportional to
the area occupied by partons. Actually, such a view on the inelastic
cross-section was suggested in the KLN approach \cite{KLN} but for
nucleus-nucleus and hadron-nucleus collision. Therefore, we generalize
this approach to hadron-hadron scattering. We believe that if the LHC
data at higher energy will support this idea, it will be a strong
argument in favor of the saturation approach. The relation 
$\sigma_{NSD} =
\sigma_{tot} - \sigma_{el} -\sigma_{sd} - \sigma_{dd}$ shows the obvious fact that the prediction for elastic and diffractive
scattering are much more complicated and less transparent in the
saturation approach. This is well-known fact at least for diffractive
production \cite{KLDD}.

The average transverse momentum of the mini-jet is defined in the usual way:
\beq   \label{PO6}
\langle p_{\mbox{jet},T }\rangle\,\,=\,\,\int d \eta \,h[\eta]\int d^2 p_T \,|p_T|\,\frac{d \sigma}{d \eta \,d^2 p_{T}}\left[\eq{MF1}\right]\Big{/} \int d \eta \,h[\eta]\int d^2 p_T \,\frac{d \sigma}{d \eta \,d^2 p_{T}}\left[\eq{MF1}\right].
\eeq
The advantage of this quantity is that it can be calculated  without usual uncertainties associated with the soft interaction physics. 
The average transverse momentum of the jet can be directly related
to the saturation scale via Eqs.~(\ref{MF1},\ref{CA5},\ref{PO6}) and it has the following simple form
at large $Q_s \gg m_{jet}$,  
\beq \label{PO7}
\langle p_{\mbox{jet},T } \rangle\,\,\propto\,\,\frac{Q_s}{\ln\Lb Q^2_s/m^2_{jet}+1\Rb+\mathcal{Q}},
\eeq
where the parameter $\mathcal{Q}$ is of order of one and takes into
account the contribution of integrals in \eq{PO6} for $p_T > Q_s$.

In order to calculate the transverse momentum of hadrons which is measured experimentally,
we need to recall that $\vec{p}_{\mbox{hadron}, T}\,\,=\,\,
z\,\vec{p}_{\mbox{jet}, T}\,+\,\vec{p}_{\mbox{intrinsic},T}$
which leads to
\beq \label{PO8}
\langle p_{\mbox{hadron},T} \rangle\,\,=\,\,\sqrt{ \,\langle z p_{\mbox{jet},T }\rangle^2\,\,+\,\,\langle p_{\mbox{intrinsic},T}\rangle^2},
\eeq
where $z$ is the fraction of energy of the jet carried by the hadron.
$\langle p_{\mbox{intrinsic},T}\rangle$ is the average intrinsic
transverse momentum of the hadron in the mini-jet. In other words, this is the
transverse momentum of the hadron in the mini-jet that has only longitudinal momentum.

In the framework of the LHPD, the $p_T$ spectrum of the produced hadron is equal to
\beq \label{PO9}
\frac{d N_{\mbox{hadron}}}{d^2 p_T}\,\,=\,\,\int d \eta\,h[\eta]\frac{1}{\sigma_{nsd}}\, \,\frac{d \sigma}{d \eta \,d^2 p_{\mbox{jet},T}}\left[\eq{MF1}~ \mbox{with}\,\, p_{ \mbox{jet}, T} \,= \,p_T/z \right],
\eeq
where in the above $p_T$ is the transverse momentum of the produced hadron.

In the CGC scenario, the gluon saturation
scale is proportional to the density of partons (see
Refs. \cite{KLN,KLNLHC}). The parton density is proportional to the
multiplicity and, therefore, we can use the following expression for
the saturation momentum in the event with the multiplicity of the
hadrons $n$:
\beq \label{PO10}
Q_s(x)\to Q_s\Lb n; x\Rb\,\,=\,\,\frac{n}{\langle n \rangle}\,Q_s\Lb x \Rb,
\eeq
where $\langle n \rangle$ is the average multiplicity that has been
measured in inclusive production without any selection related to
multiplicity  and $Q_s(x)$ is the saturation scale for inclusive hadron production or $Q_s(n=<n>;x)$.
Using \eq{PO7} again one can relate the saturation
scale at a given multiplicity to the average transverse momentum of
the produced mini-jets at large $Q_s \gg m_{jet}$,  
\beq \label{PO11}
\langle p_{\mbox{jet},T };n \rangle\,\,\propto\,\,\frac{Q_s\Lb n; x\Rb}{\ln\Lb Q_s^2\Lb n; x\Rb/m^2_{jet}+1\Rb + \mathcal{Q}}.
\eeq

\begin{figure}[!t]
       \includegraphics[width=8 cm] {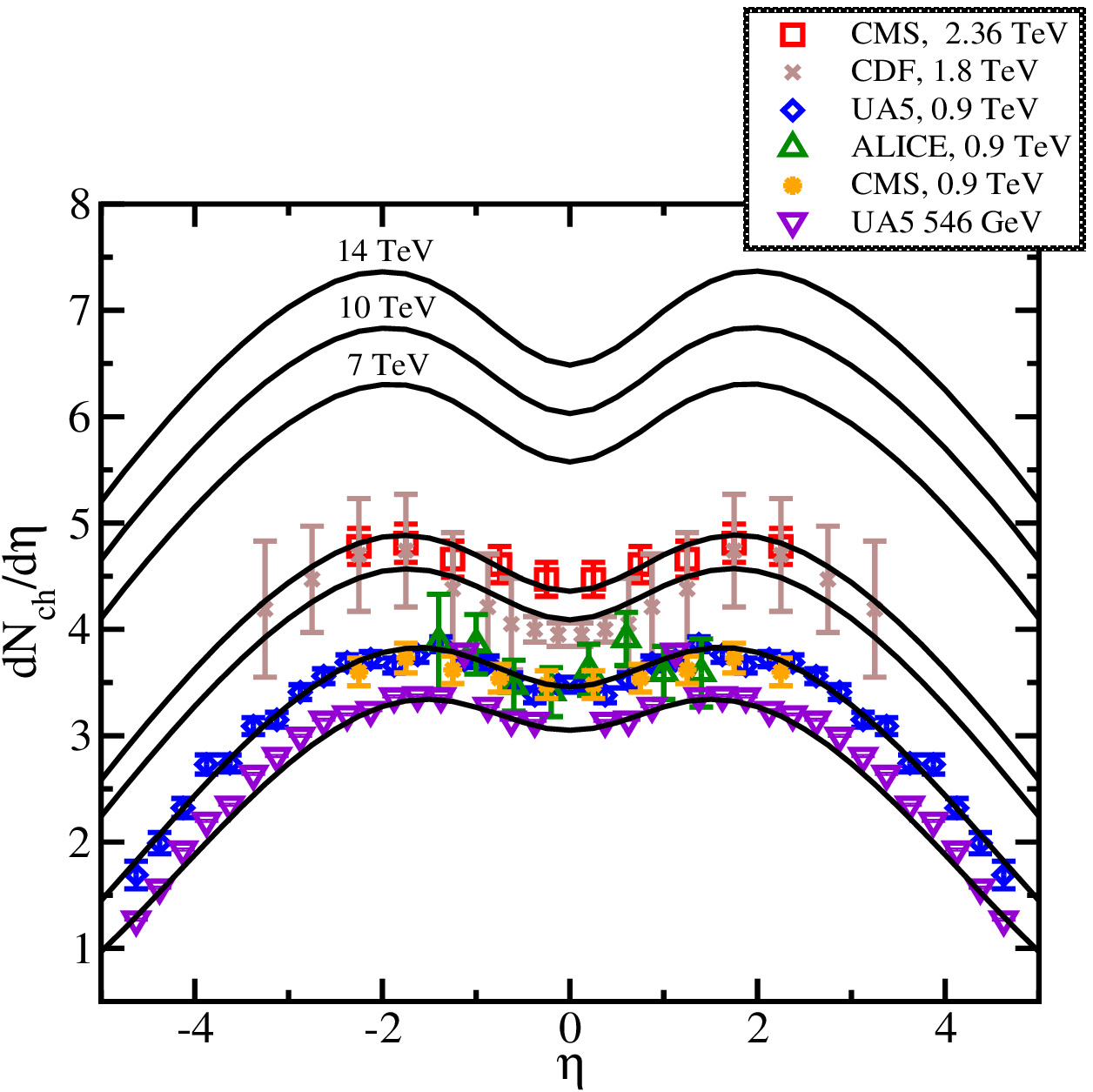}
       \includegraphics[width=7 cm] {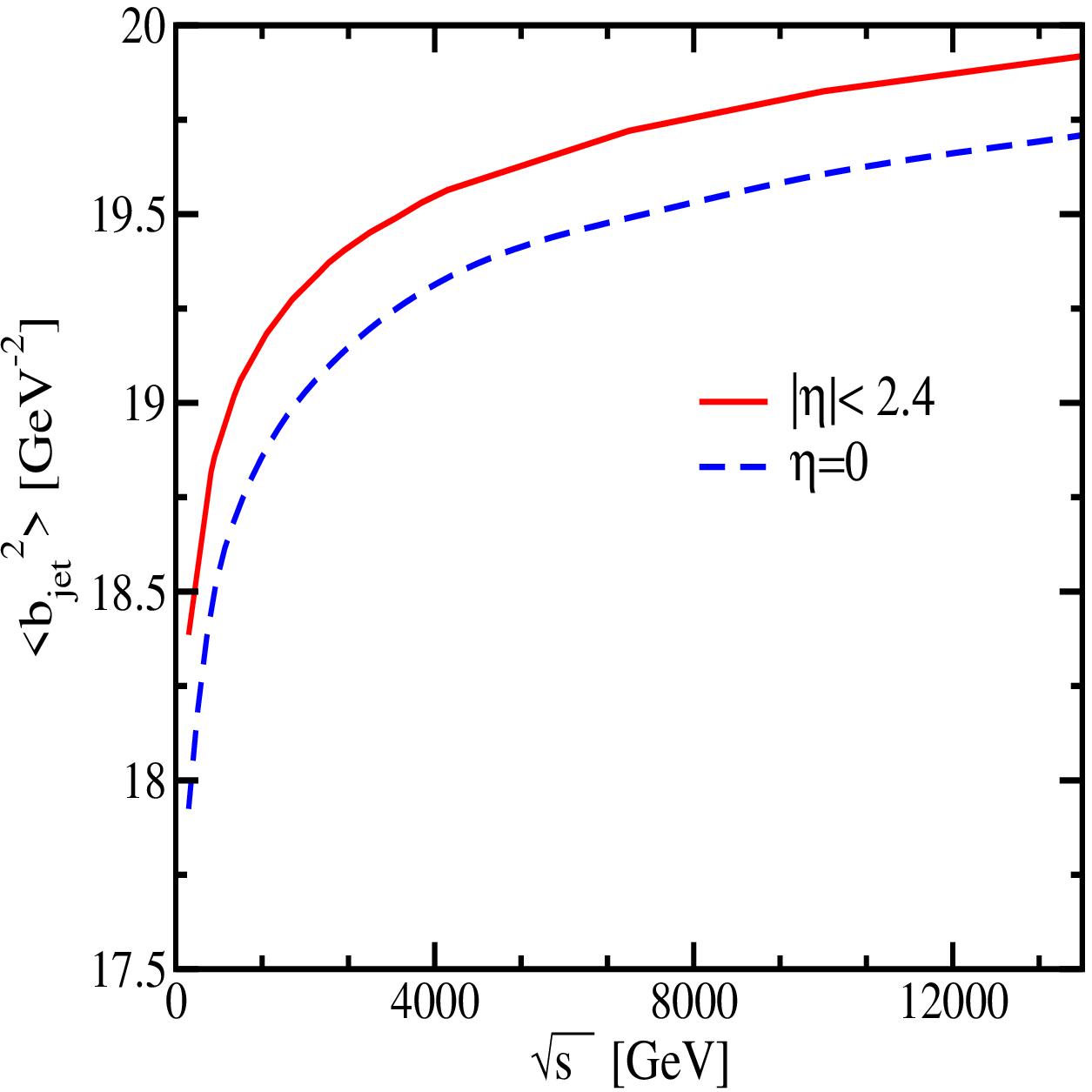} \caption{Right:
       shows the average impact parameter of the produced mini-jet
       $\langle b^2_{jet} \rangle$ given by \eq{PO5} as function of
       energy. Left: The comparison with the experimental data and
       prediction for $dN_{ch} / d y$ using \eq{PO1} with
       $\sigma_{nsd} = M\pi \langle b^2_{jet} \rangle$ for $|\eta|<2.4$.  The curves
       are normalized by data at $\sqrt{s} = 546\,\text{GeV}$, see the text
       for the details. The experimental data are from
       Refs.~\cite{AL1,CMS,particleb}. The error bars on the UA5 and
       ALICE data points are statistical. We show only
       systematic errors for the CMS data points. }
\label{f1}
\end{figure}

\section{Comparison with the experimental data and prediction for higher energies}

In the derivation of the $k_t$-factorization it was assumed that the
strong coupling $\alpha_s$ is a constant. As a generalization, 
in \eq{MF1} we replace $\alpha_s$ by $\alpha_s(p_T)$, where $p_T$ is the
transverse momentum of the mini-jet and in \eq{MF2} we also replace $1/\alpha_s$
by $1/\alpha_s(Q_s(x_i))$ where $Q_s(x_i)$ is the saturation scale in
hadron $h_i$. 
This seems to be the most natural way of introducing the running coupling which still  
preserves the form of \eq{MF4} apart from the over-all factor outside of integrals which
 now depends on kinematics. Indeed the inclusion of running strong-coupling leads to improvement of our description.  
For the running strong coupling
$\alpha_s$, we employ the same scheme as used by the KLN approach \cite{KLN},
namely we use the leading-order running coupling with smooth freezing
below the virtuality $Q^2\approx 0.8~\text{GeV}^2$ at the value of
$\alpha^{IR}_{s}\approx 0.5$. This is in accordance with many evidence from jet physics which indicates that the QCD coupling may stay
reasonably small, $\alpha^{IR}_{s}=0.4\div 0.6$ in the infrared region \cite{jet-a}.

The impact-parameter dependence in our formulation emerges from the
employed impact-parameter dependent saturation scale, see Eqs.~(\ref{CA5},\ref{CA6}).  In this model, the profile of the saturation scale in the
proton is assumed to be a Gaussian. It is difficult to interpret the
parameter $B_{CGC}$ in \eq{CA6} in terms of proton size due to the dipole size $r$ and
rapidity $Y$ dependence of the anomalous dimension. Nevertheless, in order to
have a intuitive picture, one may take $2B_{CGC}$ as relative average
squared transverse radius of the proton. The value of
$B_{CGC}=7.5~\text{GeV}^{-2}$ was obtained as a fit in order to
describe the slope of $t$-distribution of diffractive processes at
HERA \cite{WAKO}, which in turn fix the normalization of the color dipole-proton
cross-section. In Fig.~\ref{f1} (right), we show the average impact parameter of jet
$\Big \langle \vec b^2_{jet} \Big\rangle$ from center of the hadrons. Notice that for obtaining $\Big \langle \vec b^2_{jet} \Big\rangle$,
the over-all coefficient in \eq{PO5} will be dropped out and we are left
with no free parameter.  The $\Big \langle \vec
b^2_{jet} \Big\rangle$ is about $2.5 B_{CGC}$ and it slightly increases with energy. 

\begin{figure}[!t]
       \includegraphics[width=8.1 cm] {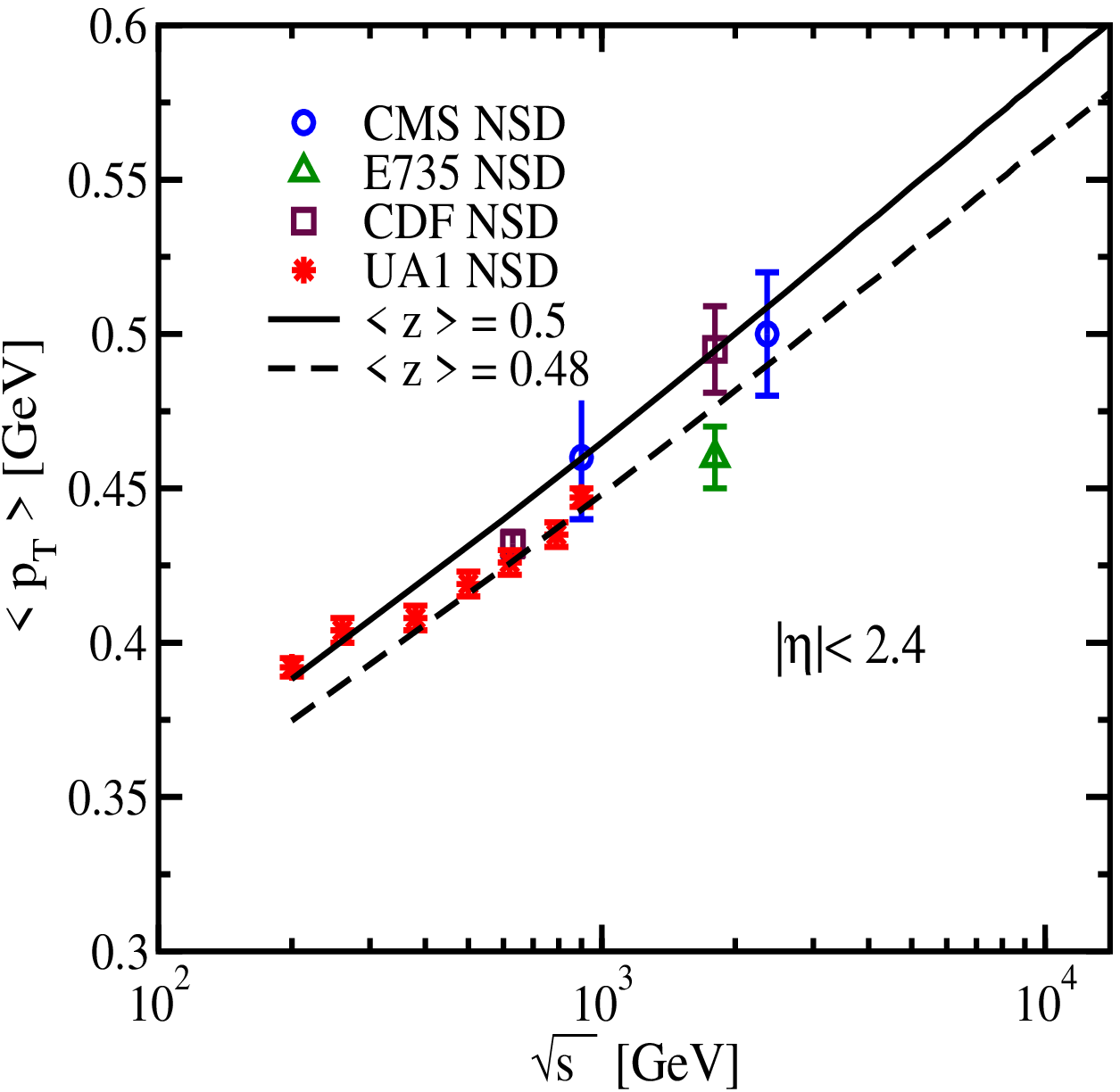}
       \includegraphics[width=8 cm] {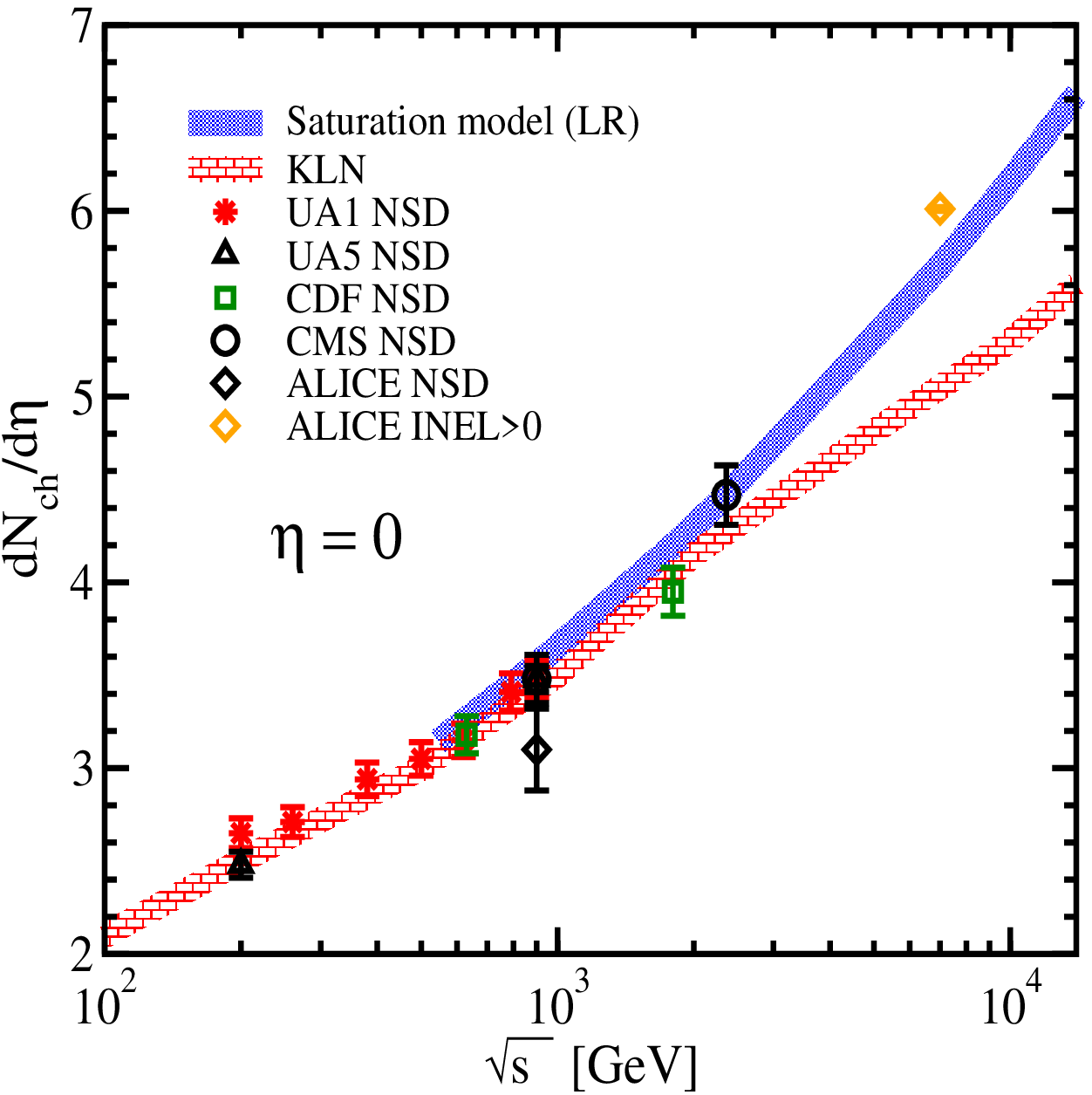}
       \caption{Right: Energy dependence of the charged hadrons
       multiplicity in the central region of rapidity $\eta=0$ in $pp$
       collisions. The theoretical curve (Saturation model LR) is our prediction coming from
       the saturation model for the NSD interactions. The band indicates
       about $2\%$ theoretical error. The total theoretical uncertainties is less $6\%$ at high energies (see the text for the details). 
       We also show the KLN prediction \cite{KLN} with the same error band as ours.
       Left: Our prediction for the energy dependence of the
       average transverse momentum of charged hadrons. The CMS data
       \cite{CMS} points and the theoretical curves in the left panel
       are for $|\eta|<2.4$. The experimental data are from
       Refs.~\cite{CMS,ALICE,particleb,ua1,ua5,CDF1,ecr}. The
       experimental error bars indicate systematic uncertainties.  }
\label{f2}           
\end{figure}

The mass of mini-jet $m_{jet}$ is proportional to the saturation scale
$m_{jet}^2 \simeq 2 \mu p_T$ \cite{KLN} since the typical transverse momentum
of the mini-jets is the saturation scale $Q_s$ and $\mu$ is the scale
of soft interaction. The saturation scale in the CGC-b model Eq.~(\ref{CA5}) changes
slowly with energy. For our interested range of energy considered in
this paper at midrapidity $\eta=0$ and $p_T=1$ GeV for the central
collisions $b=0$, we have $Q_s\approx 0.6\div 0.8$ GeV. Taking the
scale of soft interaction equal to pion mass $\mu\approx m_\pi=0.14$
GeV, we have $m_{jet}\approx 0.4\div 0.5$ GeV.  We will first assume a fixed
value for the mini-jet mass $m_{jet}=0.4$ GeV. To estimate the effect of
the mini-jet mass, we will later consider a case with a different value
for $m_{jet}$.

In order to obtain the multiplicity distribution of hadrons in $pp$
collisions from the corresponding mini-jets production cross-section Eqs.~(\ref{MF1},\ref{PO1}) we have to fix
some unknown parameters. First, based on the gluon-hadron duality, the
rapidity distribution of hadron and radiated mini-jets can be different
by a factor $C$. Second, although the $k_t$-factorization incorporates
the small-$x$ evolution taking into account the higher-order gluon
scatterings and non-linear gluon recombination effects, nevertheless given that we resort to a phenomenological color-dipole model,  
there might be still some extra contributions which are missed in our
formulation.  The discrepancy between the exact calculation and our
formulation can be then effectively taken into account with a extra $K$-factor. Finally, in order to obtain the
charged-particle multiplicity, we should divide the mini-jet
cross-section with non-singlet diffractive cross-section which as we
already discussed is obtained via $\sigma_{nsd}=M \pi \Big \langle
\vec b^2_{jet}
\Big\rangle$ with a new unknown dimensionless parameter $M$. Fortunately, these three unknown pre-factors $C, K $ and
$M$ appear as a product and can be reduced to only one unknown
parameter which will be determined with a fit to the experimental data
for the charged particle multiplicity $\frac{dN_{ch}}{d\eta}$ at
midrapidity for the lowest energy considered here $\sqrt{s}=546$
GeV. Therefore, we obtain $\frac{KC}{M}=2.32$ at $\sqrt{s}=546$
GeV. We assume that this over-all normalization factor is
energy-independent. We expect that the
energy-dependence of the normalization factor to be proportional to $1 + O(1/ln(1/x))$. Then for higher energy $\sqrt{s}>546$ GeV, we do
not have any free parameters in our calculation and our results may be
considered as predictions of the model. Notice that we have employed a
color-dipole model that its free parameters was obtained from a fit to the HERA data for $x_{B}<0.01$
and $Q^2\in[0.25, 45]$, therefore our formulation is less reliable at
lower energies (now used here). In Fig.~\ref{f1} (left), we show the charged multiplicity
distribution for $pp$ collisions at various energies. Our model gives
a good description of all available data for $\sqrt{s}\geq 546$ GeV
including the recently released data from ALICE \cite{AL1}, CMS \cite{CMS} and ATLAS \cite{ATLAS} at
$0.9$ and $2.36$ TeV. We also show our predictions for the LHC energies
at $7, 10 $ and $14$ TeV. It is seen that as the energy increases the
peak of rapidity distribution at forward (backward) becomes more
pronounced. This effect has been also observed in Ref.~\cite{mme} where it was
shown that the rapidity dependence of the invariant cross-section for
both identified hadrons and direct photon has a peak at forward
rapidities and this peak will be further enhanced by saturation
effects \cite{mme}.

\begin{figure}[!t]
              \includegraphics[width=8 cm] {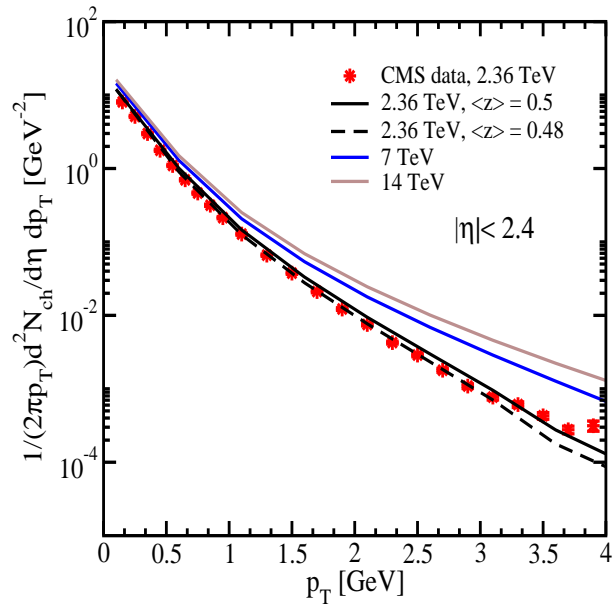} \caption{The
              differential yield of charged hadrons for
              $|\eta|<2.4$. The experimental data are from CMS
              \cite{CMS} at $2.36$ TeV for $|\eta|<2.4$. We show also
              our theoretical predictions for $7$ and $14$ TeV with
              $<z>=0.5$ and $m_{jet}=0.4$ GeV. The experimental error bars shown are systematic and
              statistical errors added linearly. }
\label{f3}         
\end{figure}

\begin{figure}[!t]
       \includegraphics[width=15 cm, height=10 cm] {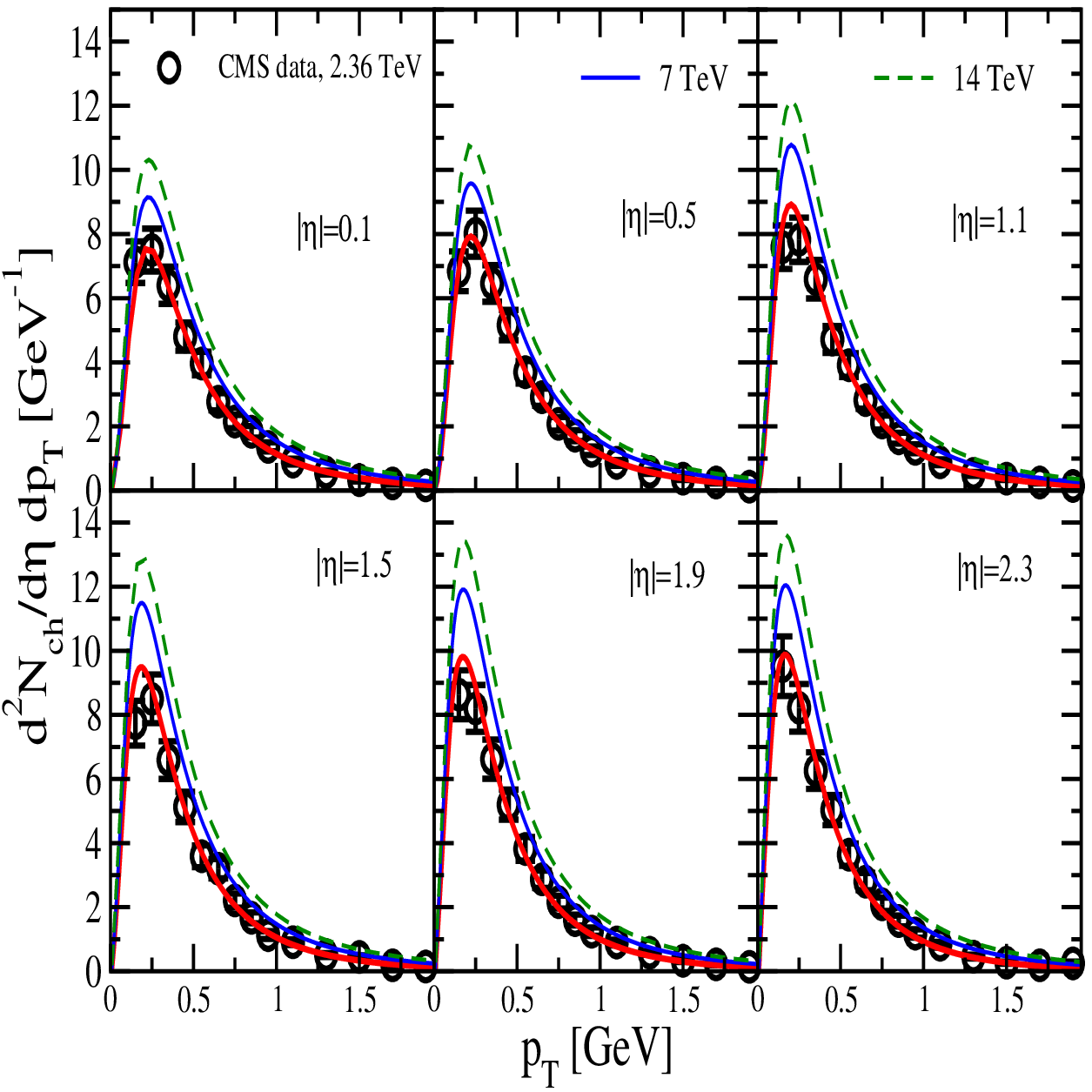}
       \includegraphics[width=8 cm] {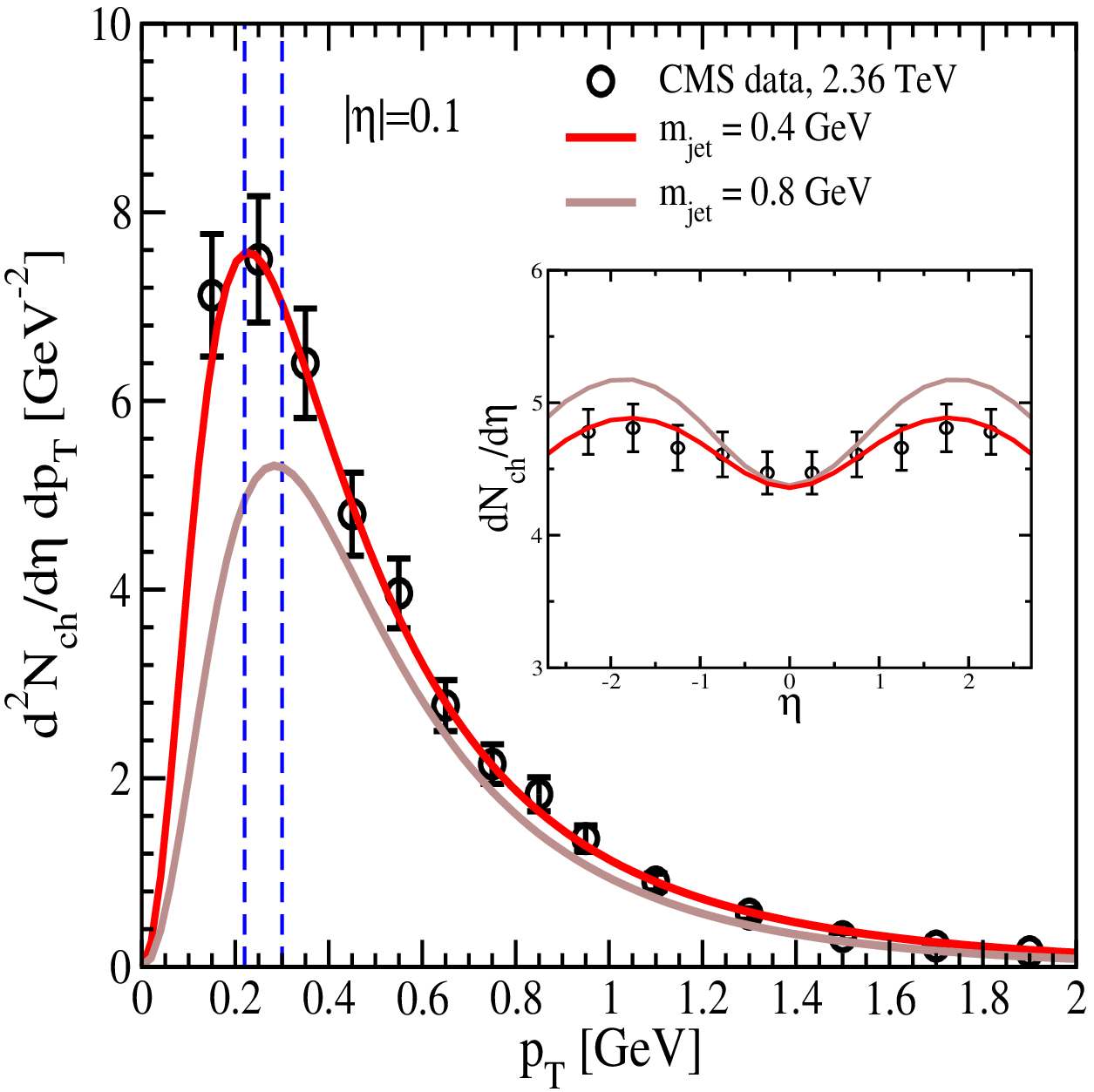} \caption{Upper
       panel: The differential yield of charged hadrons in various
       $|\eta|$ bins for $\sqrt{s}=2.36$ TeV. The experimental data
       are from CMS \cite{CMS}. We also show our
       predictions for $7$ TeV and $14$ TeV with
              $<z>=0.5$ and $m_{jet}=0.4$ GeV . The experimental error bars shown are systematic
       and statistical errors added linearly. Lower panel: The differential yield of
       charged hadrons for $|\eta|=0.1$ for two different value of
       mini-jet masses $m_{jet}$. The inserted plot in the lower panel
       figure shows the charged hadrons multiplicity again for two
       values of $m_{jet}$ for the same energy. }
\label{f4}           
\end{figure}

\begin{figure}[!t]
       \includegraphics[width=10 cm] {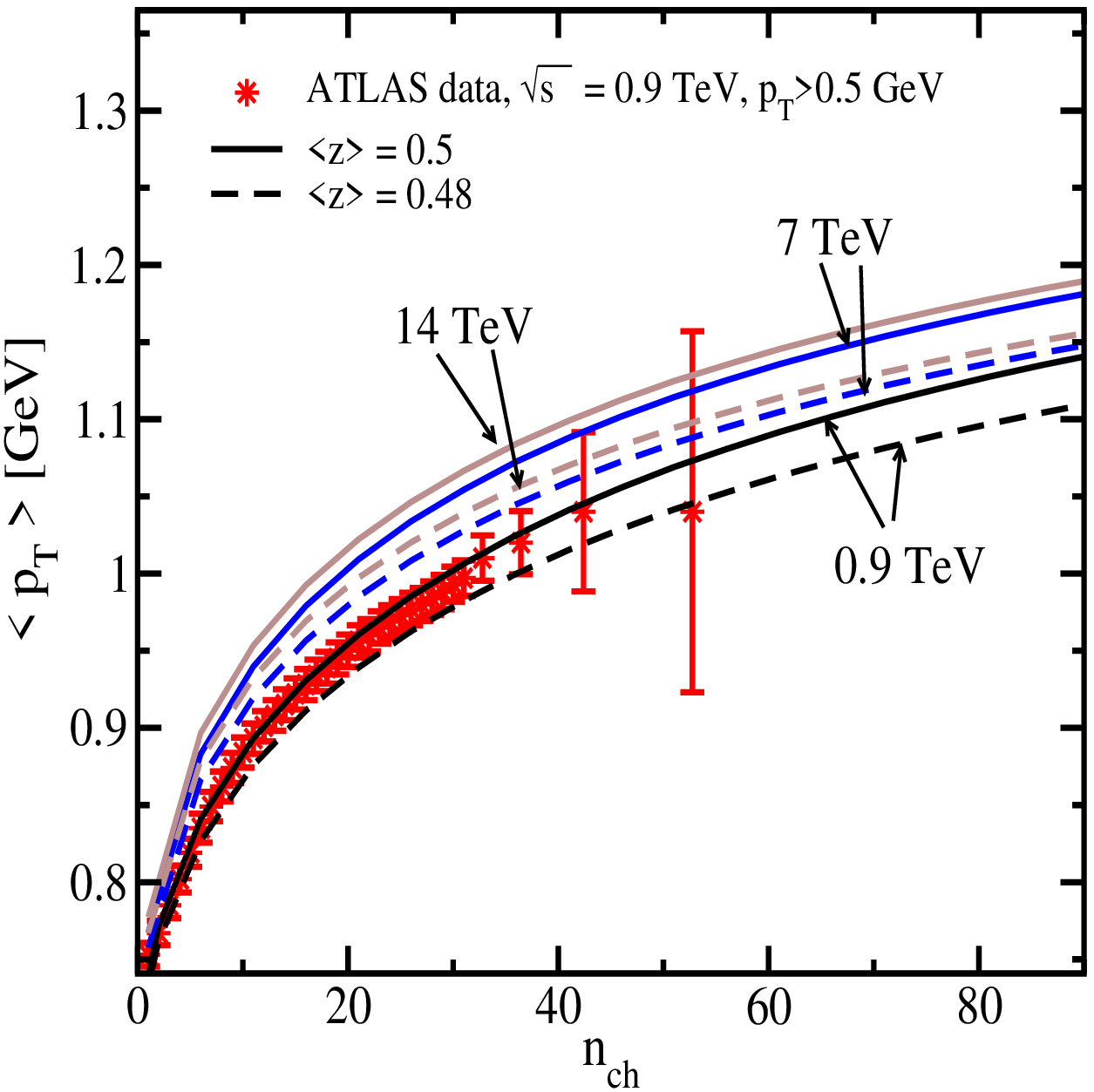}
\caption{The average transverse momentum of charged hadrons as a function of the number of charged particles for 
events with $n_{ch}\geq 1$ within the kinematic range
$p_T>500~\text{MeV}$. The experimental data are from ATLAS for
$\sqrt{s}=0.9$ TeV and $|\eta|<2.5$ \cite{ATLAS}. The theoretical curves was
obtained for $|\eta|=0$ and with the same kinematic constraint $p_T>500~\text{MeV}$ at various energies for two value of $\langle z\rangle =0.48, 0.5$
corresponding to the dashed and the solid lines, respectively. We only show the systematic experimental
errors. }
\label{f5}
\end{figure}

In Fig.~\ref{f2} (right) we show the charged-hadron pseudorapidity
density in the central region $\eta=0$ as a function of center-of-mass
energy in $pp$ collisions. Notice that since our prescription is valid
only for the NSD interactions we do not show the corresponding data
for the inelastic event selection. We have also shown recently
reported charged-particle pseudorapidity density from ALICE
\cite{ALICE} at $7$ TeV in $|\eta|<1$ for inelastic collisions with at
least one charged particle in that region (denoted by
$\text{INEL}>0$). Again this point is out of the scope of our
calculation and we did not expect to explain it.

The main source of possible theoretical error in our calculation
are due to the uncertainties associated with assuming a fixed value
for the mini-jet mass for all energies and the uncertainty in value of
energy-independent normalization factor $KC/M$ obtained from a
fit. The value of mini-jet mass is controlled by the saturation scale
and as we already discussed, it can be $m_{jet}\leq 0.65$ GeV for our
interested range of energy here.  Notice that the saturation scale in our
model varies very slowly with energy.  The upper limit of the
theoretical uncertainty band in Fig.~\ref{f2} (right) corresponds to a higher
mini-jet mass $m_{jet}=0.5$ GeV.  The experimental systematic and
statistical errors in the data point taken for fixing the
normalization also induce uncertainty in the value of pre-factor
$KC/M$ obtained from a fit. This error is included in the band shown
in Fig.~\ref{f2} (right) and is less than the uncertainties
coming from modeling the mini-jet mass. Over all we expect less than
$6\%$ theoretical error in our calculation at higher energies.

Our approach improves saturation based (KLN approach) calculation \cite{KLN} in
several ways, including: we used a correct relation between the
unintegrated gluon-density and the forward dipole-nucleon amplitude
Eqs~(\ref{MF2},\ref{MF3}) in the $k_t$-factorization Eq~(\ref{MF1}). As it is seen this
relation is not a simple Fourier transform of the dipole-amplitude which is commonly used
in literature and also depend on the impact-parameter.  
The impact-parameter dependence in these equations is not trivial and in principle 
should not be assumed as an over-all factor. We then
employed an impact-parameter dependent saturation model which was
obtained from a fit to low Bjorken-$x$ HERA data. In this sense, we
had no freedom in modeling the saturation physics compared to the KLN
approach.  Moreover, since we have an impact-parameter formulation
here, we could calculate the average relative interaction area at
higher energies and thereby could also determine the relative increase
of the NSD cross-section. It should be recalled that in the KLN approach
the information about $\sigma_{nsd}$ was taken from the models for the
soft high-energy interactions which is alien to the saturation
approach.  In both approaches, lower energy data for pp was used to
fix the overall normalization factor. Therefore, we expect that the
discrepancies between our predictions and the KLN to be more pronounced at
higher energies. This is indeed the case as it can be seen in Fig.~\ref{f2},
it is seen that the KLN prediction underestimates the multiplicity at
higher energies.

The average transverse momentum
of charge hadrons can be obtained from Eq.~(\ref{PO8}). In
Eq.~(\ref{PO8}), the average intrinsic transverse momentum of hadron
has a purely non-perturbative origin and is due to the finite-size
effect of hadrons. We take $\langle p_{\mbox{intrinsic},T}\rangle$
equal to the pion mass, the scale of soft-interaction $\mu=m_{\pi}$ throughout this paper.  
In order to obtain the average transverse momentum of charge hadrons,
we need also to know the value of the average momentum fraction of
mini-jets carried by the hadrons $\langle z\rangle$. It is seen from
Fig.~\ref{f2} (left) that an average value of $\langle z
\rangle=0.48\div 0.5$ is remarkably able to describe the average
transverse momentum of charge hadrons in a wide range of energies.
Our theoretical curves and CMS data \cite{CMS} are for the range $|\eta|<2.4$.
One may also estimate the value of $\langle z\rangle$ from the
fragmentation functions, having in mind that the $\langle z \rangle$
for mini-jets in parton-hadron duality picture is not necessarily the
same as the corresponding average of fragmentation momentum of the
produced gluons in the parton model.  Nevertheless, employing recently
developed AKK08 fragmentation functions \cite{AKK} for charged hadrons production
from a gluon, one obtains $\langle z \rangle= 0.5$ on average over low
$p_T$ within the range of $1<p_T~[\text{GeV}]\leq 2$ (AKK`s fragmentation is
valid only for $Q>1~\text{GeV}$). In order to further test the
validity of the value $\langle z \rangle \approx 0.5$ for the mini-jets, we show
in Figs.~\ref{f3}, \ref{f4}(top panel) our predictions obtained from
Eq.~(\ref{PO9}) for the differential yield of charged hadrons in the
range $|\eta|<2.4$ and at various $|\eta|$ bins for $\sqrt{s}=2.36$
TeV. The experimental data are recently reported from CMS
collaboration \cite{CMS}. It is seen that our results is in quite good agreement
with experimental data.  We recall again that the pre-factor in
Eq.~(\ref{PO9}) is the same as what we already fixed with experimental
multiplicity data at low-energy $\sqrt{s}=546$ GeV at $\eta=0$.
Therefore, we have no free parameters in obtaining the theoretical
curves in Figs.~\ref{f3},\ref{f4} (top). In Figs.~\ref{f3},\ref{f4}
(top), we have also shown our predictions for $\sqrt{s}= 7$ and $14$
TeV. The fact that our model reasonably works at low $p_T$ (for
$\sqrt{s}= 2.36$ TeV) is due to the fact that the saturation scale is
rather large at low $p_T$, for $p_T \approx m_\pi$ we have $Q_s\approx
1$ GeV in the central rapidity region. Notice that the LPHD in the simplified form that has been used here,  
is less reliable at higher $p_T$ and one should then somehow model the fragmentation of mini-jets into hadron.

In Fig.~\ref{f4}, it is seen a peculiar peak of the charged hadrons
production rate at low $p_T \approx 0.2~\text{GeV}$. Actually the
appearance of such a peak is expected in our formulation. Notice that
from Eq.~(\ref{PO1}) the differential yield of charged hadrons has a
form $\frac{d^2N}{d\eta dp_T}\propto \frac{2\pi
p_T}{p_T^2+\langle z\rangle^2 m_{jet}^2}\mathcal{F}(x_1,x_2,p_T)$ where $\mathcal{F}$ is
an analytic function. At $p_T=0$ trivially we have $\frac{d^2N}{d\eta
dp_T}=0$, for $p_T < m_{jet}\langle z\rangle$ the spectra is a monotonically
increasing function of $p_T$ and for $p_T>m_{jet}\langle z\rangle$ it is decreasing
due to the denominator. The position of the peak is then approximately at $p_T\simeq m_{jet} \langle z\rangle \approx 0.2~\text{GeV}$
since we have $\langle z\rangle=0.5$ and
$m_{jet}=0.4~\text{GeV}$. This simple picture is consistent with the
CMS experimental data \cite{CMS} shown in Fig.~\ref{f4} (top).

In order to see more clearly the effect of the mini-jet mass
$m_{jet}$, in Fig.~\ref{f4} (down) we compare the differential yield of
charged hadrons calculated with two different values for the mini-jet
mass $m_{jet}=0.4$ and $0.8~\text{GeV}$. We also show the
multiplicity distribution in the inserted panel in Fig.~\ref{f4}. As we
already pointed out, the mass of mini-jet is controlled by the
saturation scale.  Obviously from the saturation scale in our model,
$m_{jet}=0.8$ GeV is too large. Therefore, it is not surprising that
the description of experimental data for both multiplicity and spectra
worsened for such a large mini-jet mass.  Nevertheless, it is obvious
from Fig.~\ref{f4} that the position of the peak moves to a higher
$p_T$ for a larger mini-jet mass. Note that the CMS experimental data
\cite{CMS} at $\sqrt{s}=2.36$ TeV for the average transverse momentum
of charged hadrons can be reproduced with $\langle z\rangle=0.37$ when
$m_{jet}=0.8$ GeV.  Again the position of the peak in spectra is
consistent with simple formula $p_T \simeq m_{jet}
\langle z\rangle \approx 0.3$ in accordance with the full calculation
shown in Fig.~\ref{f4}.  Notice that in our model calculation shown in
Fig.~\ref{f4} (top), the position of the peak persists at various
rapidities bin (and energies) since we have taken a fixed $m_{jet}$
for simplification. To conclude, a precise measurement of the
differential yield of charged hadrons at low $p_T$ for higher energies
at LHC will provide valuable information about the mini-jet mass
and its connection with the gluon saturation. 

In Fig.~\ref{f4} (top), we also showed our theoretical predictions for $7$ and $14$ TeV
with a fixed $\langle z\rangle=0.5$ and $m_{jet}=0.4$ GeV. As we
already explained due to the possible increase of mini-jet mass at higher energies, the
position of peak may slightly move to higher $p_T$ within $0.2\le p_T
[\text{GeV}] \leq 0.3$ at $\sqrt{s}=14 $ TeV.

In Fig.~\ref{f5}, we show the average transverse momentum of charged
hadrons as a function of the number of charged particles for events
within the kinematic range $p_T>500~\text{MeV}$. The experimental data
are from ATLAS for $\sqrt{s}=0.9$ TeV \cite{ATLAS}. The saturation scale at various multiplicity is given
by Eq.~(\ref{PO10}) where $\langle n\rangle$ can be conceived as a
normalization and its value is taken to be the charged multiplicity at
midrapidity $\eta=0$ for a given center-of-mass energy (shown in
Fig.~\ref{f2} (right)). In order to implement in our calculation the
experimental kinematic constrain $p_T>500~\text{MeV}$ on the measured
events, we impose that $\langle p_{\mbox{intrinsic},T}\rangle
>500~\text{MeV}$. The $\langle
p_{\mbox{intrinsic},T}\rangle$ has a purely non-perturbative origin
and can be of order of hadron mass. To this end, we take $\langle
p_{\mbox{intrinsic},T}\rangle=(m_{\rho}+m_{k})/2$ where the mass of
$\rho$ and $k$ mesons are $m_{\rho}=775$ MeV and $m_{k}=497$ MeV,
respectively. In Fig.~\ref{f5}, we show $\langle p_T\rangle$ for two
values of $\langle z\rangle$. It is seen that our model is able to
give a very good description of the ATLAS data. We have also shown in
the same plot, our predictions for the higher LHC energies. 

The general behavior of the theoretical curves shown in Fig.~\ref{f2} (left) and Fig.~\ref{f5}
for the average transverse momentum of the produced hadrons is in
accordance with a simple formulas given in
Eqs.~(\ref{PO7},\ref{PO11}) showing a clear connection between the
gluon saturation and the measured transverse momentum of charged
hadrons.


\section{Conclusion}

In high density QCD the main source of hadron production is the decay
of gluon mini-jets with the transverse momentum of the order of the
saturation scale. This viewpoint is based on the fact that the system of
partons (gluons) creates a new state of matter, the so-called Color Glass Condensate, in which the gluon density reaches the limited values of the order
of $1/\as$ with new typical transverse momentum (the saturation scale). We
developed a model that includes the gluon saturation and demonstrated
that this model is able to describe both the inclusive hadron
production at high energies including the first data from the LHC and
the deep inelastic scattering data from HERA in a unique fashion. 

We predicted an increase of $d N_{ch}/d y|_{\eta =0}$, mean
transverse momentum and the multiplicity of produced charged hadrons with energy which
is in accordance with the first LHC data measured by ALICE \cite{AL1, ALICE}, CMS \cite{CMS} and
ATLAS \cite{ATLAS} collaboration, see Figs.~\ref{f1},\ref{f2},\ref{f5}. In the framework of high density QCD all these phenomena are
closely related to the growth of the saturation momentum as a function
of energy and of density of partons.  It should be stressed that the other
high-energy phenomenological approaches \cite{GLMINC} cannot describe the dependence of the
average transverse momentum of the produced hadron on energy and
hadron multiplicities.

We showed that recently reported data by the CMS collaboration \cite{CMS} on
the differential yield of charged hadrons at low $p_T$ for $\sqrt{s}=2.36 $ TeV 
reveal an interesting information on the mini-jets production and its connection with the saturation.
We showed that the appearance of a peak in differential yield of charged hadrons at low $p_T$
is closely related to the mini-jet mass and the value of the saturation scale.
  
We provided various predictions for the upcoming LHC measurements at
higher energies in $pp$ collisions. We believe that this paper will be useful for the microscopic
interpretation of the upcoming LHC data and will lead to a deeper
understanding of the hadron interactions at high energy in the
framework of QCD.

Concluding this paper we would like to answer the question: what can be here considered as a possible 
signal of the saturation (CGC) which are not contaminated with the
non-perturbative physics related to unknown confinement of quarks and
gluon? The main non-perturbative parameter that we have to introduce
is $m_{jet}$. The rapidity distribution $d N_{ch}/d \eta$ at $|\eta| <
1$ (Fig.~\ref{f1} ), the $p_t$ spectrum of hadron at low $p_T \leq m_{jet}$
and the position of the maximum in $d^2 N_{ch}/d \eta d p_T$
(Fig.~\ref{f4} ) depend on the value of $m_{jet}$ and the success of our description indicates that we have chosen this parameter in
self-consistent way.  However, the energy dependence of $d N_{ch}/d
\eta$ at $|\eta|\leq 3.5$ and the average value of the transverse
momentum $<p_T>$ of hadrons as well as the multiplicity dependence
of $<p_T>$ and the rapidity dependence of the maximum in $d^2 N_{ch}/d
\eta d p_T$ for $|\eta|\leq 3.5$ are the typical consequences of the
saturation approach since the main contribution in the calculations
of these observables is originated from the transverse momenta of the
order of $Q_s$.  Two factors determine the behavior of the
observables at $ |\eta|\geq 3.5$ : $( 1 - x)^4 $ suppression of the
gluon densities in projectile and the increase of the saturation
momentum in the target. Since the $( 1 - x)^4 $ factor reflects the well-known
behavior of the structure function $F_2$ at large-$x$, this factor will be the same in all
other approaches while the additional increase due to the energy dependence is a typical features of the saturation approach.
 Notice also that at LHC energy $\sqrt{s}=14$ TeV, the contribution of $( 1 - x)^4 $ correction of unintegrated gluon density within the rapidities region
considered here (Fig. \ref{f1}) is negligible and at $7$ TeV this contribution is less than $5\%$.

The above discussion shows that the comparison of our prediction with
the high LHC energy data will be crucial for our approach.  We are
happy to make predictions before the experimental data from the LHC at
high energy. We believe that if the coming data confirms our
predictions, this will be indeed a first important step toward discovery of the CGC phase of the matter
at LHC. The fact that we had to introduce
several phenomenological parameters reflects our lack of
theoretical knowledge for quark and gluon
confinement and cannot be overcome in any models. Our experience tells
us that when the data for higher energies will be published a lot of
phenomenological models will appear but the CGC (saturation) approach is
the only one that gives the predictions. It has happened once for
nucleus-nucleus scattering at RHIC and, we hope that the situation
will repeat itself at the LHC.

The particle production scheme presented in this paper can be also
applied to the calculation of inclusive hadron production in heavy ion collisions at
LHC. We are currently working on this problem and plan to report on this in the near future.

\section*{Acknowledgments} 
We are thankful to Yuri Kovchegov for drawing our attention to Ref.~\cite{yuri-b}. This work was supported in part by
Conicyt Programa Bicentenario PSD-91-2006 and the Fondecyt (Chile)
grants 1090312 and 1100648.

\end{document}